\documentclass[aps,10pt,prb,onecolumn,floatfix]{revtex4-1}
\usepackage{amssymb,amsmath,bm,graphics,epsfig,mathrsfs}
\usepackage{color}

\newcommand{\mac}{\mathcal}
\newcommand{\msc}{\mathscr}
\newcommand{\tx}{\rm}
\newcommand{\ti}{\textit}

\newcommand{\dn}{\downarrow}

\newcommand{\nn}{\nonumber}
\newcommand{\pat}{\partial}
\newcommand{\para}{\parallel}
\newcommand{\pr}{\prime}

\newcommand{\raw}{\rightarrow}
\newcommand{\til}{\tilde}

\newcommand{\up}{\uparrow}

\newcommand{\ub}{\underbrace}

\newcommand{\alp}{\alpha}
\newcommand{\dlt}{\delta}
\newcommand{\Dlt}{\Delta}

\newcommand{\eps}{\epsilon}
\newcommand{\gm}{\gamma}
\newcommand{\Gm}{\Gamma}

\newcommand{\vp}{\varphi}

\newcommand{\vs}{\varsigma}

\newcommand{\og}{\omega}
\newcommand{\Og}{\Omega}
\newcommand{\sg}{\sigma}
\newcommand{\Sg}{\Sigma}

\newcommand{\ie}{{i.e.,~}}

\begin{document}
\title{Spin-orbit coupled transport and spin torque
in a ferromagnetic heterostructure}
\author{Xuhui Wang}
\email{xuhui.wang@kaust.edu.sa}
\affiliation{King Abdullah University of Science and Technology
(KAUST), Physical Science and Engineering Division, Thuwal
23955-6900, Saudi Arabia}
\author{Christian Ortiz Pauyac}
\affiliation{King Abdullah University of Science and Technology
(KAUST), Physical Science and Engineering Division, Thuwal
23955-6900, Saudi Arabia}
\author{Aur\'{e}lien Manchon}
\email{aurelien.manchon@kaust.edu.sa} \affiliation{King Abdullah
University of Science and Technology (KAUST), Physical Science and
Engineering Division, Thuwal 23955-6900, Saudi Arabia}

\begin{abstract}
Ferromagnetic heterostructures provide an ideal platform to
explore the nature of spin-orbit torques arising from the
interplay mediated by itinerant electrons between a Rashba-type
spin-orbit coupling and a ferromagnetic exchange interaction. For
such a prototypic system, we develop a set of coupled diffusion
equations to describe the diffusive spin dynamics and spin-orbit
torques. We characterize the spin torque and its two
prominent--\ti{out-of-plane} and \ti{in-plane}--components for a
wide range of relative strength between the Rashba coupling and
ferromagnetic exchange. The symmetry and angular dependence of the
spin torque emerging from our simple Rashba model is in an
agreement with experiments. The spin diffusion equation can be
generalized to incorporate dynamic effect such as spin pumping and
magnetic damping.
\end{abstract}
\maketitle

\section{Introduction}
Spin-orbit coupling is a key mechanism in many prominent physical
phenomena ranging from the electrically generated bulk spin
polarization,\cite{dp-she-1971,edelstein-1989} to the
dissipationless spin current in bulk semiconductors,
\cite{murakami-science-2004} to the spin-Hall effect in metals
\cite{she} and two-dimensional electron gas.
\cite{sinova-prl-2004,kato-science-2004} In searching for an
efficient mechanism for magnetization switching, interplay between
spin-orbit coupling and magnetism\cite{nunez-ssc-2006,tatara} has
brought a new member to the spin-transfer torque community,
\cite{slonczewski-berger-1996, refstt} i.e., the spin-orbit
torque. Diluted magnetic semiconductors provide an ideal platform
to theoretically\cite{dms-torque-theory} and experimentally
\cite{chernyshov,fang,endo} study the current-driven magnetization
dynamics induced by spin-orbit torques. Simply speaking,
spin-orbit torque operates through the competition between the
exchange and spin-orbit fields in polarizing the itinerant
electrons (or holes) and gives rise to a torque on the
ferromagnetic order parameter. When it comes to magnetization
switching, the advantage of spin-orbit torque over the
conventional one is clear: there is no need to employ a separate
ferromagnet as polarizer.

Recent experiments on current-driven magnetization dynamics
performed in multilayer systems
\cite{mihai1,pi,suzuki,mihai2,miron-nature-2011} have also
achieved current-induced switching in a \ti{single} ferromagnet
film sandwiched between a heavy metal and metal oxide, which
indicates the presence of spin-orbit coupling and therefore
spin-orbit torque as a potential driving force. These
systems--mostly consisting of conducting interfaces between
ferromagnetic metal films and heavy metals (or metal oxides)--are
nowadays often referred to as ferromagnetic heterostructures. A
viable candidate believed to exist in such structures is the
Rashba-type spin-orbit interaction due to inversion symmetry
breaking. \cite{rashba-soi} Theoretical efforts are made to
uncover dominant components of the spin-orbit torque induced by
the Rashba coupling. \cite{manchon-prb,others,wang-manchon-2011,
pesin-prb-2012,vanderbijl-prb-2012} They usually treat the
coexistence of ferromagnetism and spin-orbit coupling as an
intrinsic property. The underlying physics is fairly simple and
intuitive: when a charge current is applied in the structure, the
Rashba spin-orbit coupling creates an effective magnetic field
(coined as the Rashba field $\bm{B}_{R}$); so long as $\bm{B}_{R}$
polarizes charge carriers to the direction that is misaligned with
the magnetization direction $\bm{m}$, a spin torque emerges to act
on the magnetization and to induce switching. This torque, named
as Rashba spin-orbit torque or Rashba torque, has gained much
attention from academia as well as industries and is exactly the
central topic of this paper. Both theories and experiments have
shown that the Rashba torque shall, in general, comprise two major
components, i.e. a fieldlike torque (or in-plane component) and a
damping-like torque (or out-of-plane component). More recently,
the symmetry of spin-orbit torque has been scrutinized
experimentally. The experiments by Garello \ti{et al} reveal an
intriguing yet complex angular dependence on the magnetization
direction.\cite{garello} This observation challenges the commonly
accepted form described by in-plane and out-of-plane components.

In this paper, we provide a systematic theoretical study on the
spin-orbit torque and spin dynamics in a ferromagnetic ultrathin
film without structure inversion symmetry. We construct a simple
two-dimensional model that accommodates both a Rashba spin-orbit
coupling and an exchange interaction. For the Rashba torque, we
propose a general form that not only contains the in-plane and
out-of-plane components but also possesses symmetry and complex
angular dependence supported by experiments. In Sec.
\ref{sec:diffusion-equation}, we employ the quantum kinetic
equation to derive coupled diffusion equations for the charge and
spin densities. We account for the fact that Rashba coupling not
only produces an effective magnetic field but also induces spin
relaxation through the D'yakonov-Perel mechanism \cite{dp} that is
dominant in a quasi-two-dimensional system. In the absence of
magnetism, analytical and numerical solutions in Sec.
\ref{sec:she} are able to describe the spin-galvanic and spin-Hall
effects. We demonstrate in Sec. \ref{sec:diff-ferro} that the
diffusion equation provides a coherent framework to describe the
spin dynamics in a ferromagnetic metal. Section
\ref{sec:rashba-spin-torque} employs the spin diffusion equation
given in Sec. \ref{sec:diffusion-equation} to analyze the general
symmetry properties and angular dependence of the Rashba torque in
the limits of both a weak and strong spin-orbit coupling. We are
able to provide for the spin-orbit torque an angular dependence
that agrees well with recent experiments. In Sec.
\ref{sec:numeric}, we evaluate the spin density and Rashba torque
numerically for a wide range of relative strength between the
Rashba coupling and exchange splitting. In Sec.
\ref{sec:dynamics}, we further show the formulation proposed in
Sec. \ref{sec:diffusion-equation} can be generalized to describe
spin pumping and magnetic damping by an inclusion of temporal and
spatial variations of ferromagnetic order parameter. We find
agreement with earlier results approached using other methods.
Section \ref{sec:discussion} discusses the validity of the Rashba
model and outlines a brief comparative study between the Rashba
torque and spin-Hall effect-induced torque. Section
\ref{sec:conclusion} concludes the article.

Nevertheless, we emphasize that we make no attempt to argue that
the Rashba model provides the ultimate answer to the spin-orbit
torque in ferromagnetic heterostructures. Spin-Hall effect
\cite{she} in the nonmagnetic metal layer provides an alternative
explanation to several experiments.\cite{liu} However, we must
admit, as Haney \ti{et al.} have pointed out, both explanations
have their strength and weakness.\cite{haney-prb-2013} Despite the
limitations, the results presented here and their agreement with
experiments lead us to believe such a simple model does shed light
on the nature of the spin-orbit torque in ferromagnetic thin
films.

\section{From Hamiltonian to diffusion equation}
\label{sec:diffusion-equation} 
\begin{figure}
\centering
\includegraphics[trim = 50mm 120mm 40mm 120mm, clip, scale=0.9]{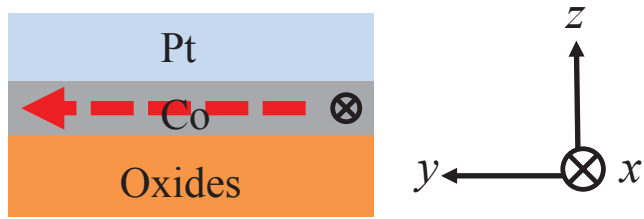}
\caption{\label{fig:setup} (Color online) Schematic view of the
cross section of a typical ferromagnetic heterostructure that
accommodates both a Rashba spin-orbit coupling and an exchange
interaction. An ultrathin ferromagnetic metal film (e.g., Co) is
sandwiched between an oxide (e.g., AlO$_{x}$) and a heavy metal
layer (e.g., Pt or Ta). A charge current is injected into the
ferromagnetic layer along the $\hat{\bm{x}}$ direction. The dashed
red arrow points to the direction of the effective Rashba field.
We shall note that the system is not isolated but connected to an
external source and drain.}
\end{figure}

In Fig. \ref{fig:setup}, we sketch
a schematic view of a cross-section of a typical ferromagnetic
heterostructure under investigation: a ferromagnetic ultrathin
metal film (rolled out in the $x$-$y$ plane) is sandwiched by a
heavy metal layer and an oxide; two asymmetric interfaces provide
a \ti{weak} confinement in the $z$ direction, along which the
inversion symmetry is broken. The potential gradient across the
interface generates a Rashba spin-orbit coupling.\cite{rashba-soi}
Without loss of generality, our starting point is therefore a
simplified quasi-two-dimensional single-particle Hamiltonian
($\hbar=1$ is assumed throughout),
\begin{equation}
{\hat H}=\frac{\hat{\bm{k}}^{2}}{2m}+\alp
\hat{\bm{\sg}}\cdot(\hat{\bm{k}}\times\hat{\bm{z}})
+\frac{1}{2}\Dlt_{xc} \hat{\bm{\sg}}\cdot\bm{m}+\hat{H}^{i},
\label{eq:hamitonian}
\end{equation}
for an electron with momentum $\hat{\bm{k}}$. In Eq.
(\ref{eq:hamitonian}), $\hat{\bm{\sg}}$ is the Pauli matrix, $m$
the effective mass, and $\bm{m}$ the magnetization direction. The
ferromagnetic exchange splitting is given by $\Dlt_{xc}$ and
$\alp$ represents the Rashba constant (parameter). The Hamiltonian
${\hat H}^{i}=\sum_{j=1}^{N}V(\bm{r}-\bm{r}_{j})$ accounts for all
nonmagnetic impurity scattering potentials $V(\bm{r})$ localized
at $\bm{r}_{j}$. Throughout the following discussion, we assume
that the exchange interaction and spin-orbit splitting are smaller
than the Fermi energy, while leaving the ratio of the spin-orbit
coupling to the exchange interaction arbitrary.

Before we proceed to detailed discussion, we clarify the validity
of such a quasi two-dimensional model. In principle, carrier
transport in the system under consideration is a three-dimensional
phenomenon in which size effect may arise. Here, our quasi
two-dimensional Rashba model assumes a direct coupling between the
exchange and effective Rashba field and is thus requiring
ultrathin layers in which diffusive motion normal to the thin film
plane can be neglected. Haney \ti{et al} have recently conducted a
thorough discussion using a three-dimensional semiclassical
Boltzmann description in a bilayer structure augmented by an
interfacial Rashba spin-orbit coupling.\cite{haney-prb-2013} Their
results are consistent with the ones obtained from
quasi-two-dimensional transport modeled in Refs.
[\onlinecite{manchon-prb,others,wang-manchon-2011,pesin-prb-2012,vanderbijl-prb-2012,kim2012-prb}].

To derive a diffusion equation for the nonequilibrium charge and
spin densities, we apply the Keldysh formalism.
\cite{rammer-smith-rmp-1986} We use the Dyson equation, in a
two-by-two spin space, to obtain a kinetic equation that assembles
the retarded (advanced) Green's function $\hat{G}^{R}$
($\hat{G}^{A}$), the Keldysh component of the Green function
$\hat{G}^{K}$, and the self-energy $\hat{\Sg}^{K}$, \ie
\begin{equation}
[\hat{G}^{R}]^{-1}\hat{G}^{K}-\hat{G}^{K}[\hat{G}^{A}]^{-1}
=\hat{\Sg}^{K}\hat{G}^{A}-\hat{G}^{R}\hat{\Sg}^{K},
\label{eq:kinetic-equation}
\end{equation}
where all Green's functions are full functions with
interactions taken care of by the self-energies $\hat{\Sg}^{R,A,K}$.
The retarded (advanced)
Green's function in momentum and energy space is
\begin{equation}
\hat{G}^{R(A)}(\bm{k},\eps)=\frac{1}
{\eps-\eps_{\bm{k}}-\hat{\bm{\sg}}\cdot\bm{b}(\bm{k})-\hat{\Sg}^{R(A)}(\bm{k},\eps)},
\end{equation}
where $\eps_{\bm{k}}=\bm{k}^{2}/(2m)$ is the single-particle
energy. The impurity scattering has been taken into account by the
self-energy, as to be shown below. We have introduced a
$\bm{k}$-dependent total effective field
$\bm{b}(\bm{k})=\Dlt_{xc}\bm{m}/2+\alp(\bm{k}\times\hat{\bm{z}})$
with magnitude
$b_{k}=|\Dlt_{xc}\bm{m}/2+\alp(\bm{k}\times\hat{\bm{z}})|$ and
direction $\hat{\bm{b}}=\bm{b}(\bm{k})/b_{k}$.

We neglect localization effects and electron-electron interactions
and assume a short-range $\dlt$-function type impurity scattering
potential. At a low impurity concentration and a weak coupling to
electrons, a second-order Born approximation is
justified,\cite{rammer-smith-rmp-1986} \ie the self-energy due to
impurity scattering is \cite{mishchenko-prl-2004}
\begin{equation}
\hat{\Sg}^{K,R,A}(\bm{r},\bm{r}^{\pr})
=\frac{\dlt(\bm{r},\bm{r}^{\pr})}{m\tau}\hat{G}^{K,R,A}(\bm{r},\bm{r}),
\end{equation}
where the momentum relaxation time is given by
\begin{equation}
\frac{1}{\tau}\approx 2\pi\int \frac{d^{2}\bm{k}^{\pr}}{(2\pi)^{2}}
|V(\bm{k}-\bm{k}^{\pr})|^{2} \dlt(\eps_{\bm{k}}-\eps_{\bm{k}^{\pr}}).
\end{equation}
$V(\bm{k})$ is the Fourier transform of the scattering potential and
the magnitude of $\bm{k}$ and $\bm{k}^{\pr}$ is evaluated at
Fermi vector $k_{F}$.

The quasiclassical distribution function
$\hat{g}\equiv\hat{g}_{\bm{k},\eps}(T,\bm{R})$, defined as the
Wigner transform of the Keldysh function
$\hat{G}^{K}(\bm{r},t;\bm{r}^{\pr},t^{\pr})$, is obtained by
integrating out the relative spatial-temporal coordinates while
retaining the center-of-mass ones $\bm{R}=(\bm{r}+\bm{r}^{\pr})/2$
and $T=(t+t^{\pr})/2$.  The spatial profile of the quasiclassical
distribution function is considered smooth on the scale of Fermi
wavelength,\; we may thus apply the gradient expansion technique
on Eq. (\ref{eq:kinetic-equation}),\cite{rammer-book} which gives
us a transport equation for macroscopic quantities. Under the
gradient expansion, the left-hand side of Eq.
(\ref{eq:kinetic-equation}) becomes
\begin{eqnarray}
[\hat{G}^{R}]^{-1} &\hat{G}^{K}& - \hat{G}^{K}[\hat{G}^{A}]^{-1} \nn\\
& \approx & [\hat{g},\hat{\bm{\sg}}\cdot\bm{b}(\bm{k})]
+\frac{i}{\tau}\hat{g}
+i\frac{\pat\hat{g}}{\pat T}
+\frac{i}{2} \left\{\frac{\bm{k}}{m}
+\alp(\hat{\bm{z}}\times\hat{\bm{\sg}}),\nabla_{\bm{R}}\hat{g}\right\},
\end{eqnarray}
where $\{\cdot,\cdot\}$ is the anticommutator. The relaxation time
approximation renders the right-hand side of Eq.
(\ref{eq:kinetic-equation}) as
\begin{equation}
\hat{\Sg}^{K}\hat{G}^{A}  -\hat{G}^{R}\hat{\Sg}^{K} \approx
\frac{1}{\tau}\left[\hat{\rho}(\eps,T,\bm{R})\hat{G}^{A}(\bm{k},\eps)
-\hat{G}^{R}(\bm{k},\eps)\hat{\rho}(\eps,T,\bm{R})\right]
\end{equation}
where we have introduced the density matrix by
integrating out $\bm{k}^{\pr}$ in $\hat{g}$, \ie
\begin{equation}
\hat{\rho}(\eps,T,\bm{R})
=\frac{1}{2\pi \mac{N}}\int \frac{d^{2}\bm{k}^{\pr}}{(2\pi)^{2}}
\hat{g}_{\bm{k}^{\pr},\eps}(T,\bm{R}),
\end{equation}
where $\mac{N}$ is the density of states for one spin specie.

For the convenience of discussion,
the time variable is changed from $T$ to $t$.
At this stage, we have a kinetic equation depending on
$\hat{\rho}$ and $\hat{g}$
\begin{eqnarray}
i[\hat{\bm{\sg}}\cdot\bm{b}(\bm{k}),\hat{g}]
&+& \frac{1}{\tau}\hat{g} +\frac{\pat\hat{g}}{\pat
t}+\frac{1}{2}\left\{\frac{\bm{k}}{m}
+\alp(\hat{\bm{z}}\times\hat{\bm{\sg}}),\nabla_{\bm{R}}\hat{g}\right\} \nn\\
&=& \frac{i}{\tau}\left[\hat{G}^{R}(\bm{k},\eps)\hat{\rho}(\eps)
-\hat{\rho}(\eps)\hat{G}^{A}(\bm{k},\eps)\right].
\end{eqnarray}
We perform a Fourier transformation on temporal variable to the
frequency domain $\og$, which leads to
\begin{equation}
\Og\hat{g}-b_{k}[\hat{U}_{k},\hat{g}]
=i\hat{K},
\label{eq:qke-with-K}
\end{equation}
where $\Og=\og+i/\tau$ and the operator $\hat{U}_{k}\equiv
\hat{\bm{\sg}}\cdot\hat{\bm{b}}$ satisfies
$\hat{U}_{k}\hat{U}_{k}=1$. The right-hand side of Eq.
(\ref{eq:qke-with-K}) is partitioned according to
\begin{eqnarray}
\hat{K} = && \ub{\frac{i}{\tau}\left[\hat{G}^{R}(\bm{k},
\eps)\hat{\rho}(\eps)
-\hat{\rho}(\eps)\hat{G}^{A}(\bm{k},\eps)\right]}_{\hat{K}^{(0)}}\nn\\
&+& \ub{-\frac{1}{2}\left\{\frac{\bm{k}}{m}
+\alp(\hat{\bm{z}}\times\hat{\bm{\sg}}),\nabla_{\bm{R}}\hat{g}\right\}}_{\hat{K}^{(1)}},
\end{eqnarray}
where $\hat{K}^{(0)}$ contributes to the lowest-order solution to
$\hat{g}$ and the gradient correction $\hat{K}^{(1)}$ is treated
as a perturbation. Both functions $\hat{g}$ and $\hat{\rho}$ are
in the frequency domain.

Equation (\ref{eq:qke-with-K}) is solved formally to give a
solution to $\hat{g}$:
\begin{equation}
\hat{g} = i\frac{(2 b_{k}^{2}-\Og^{2})\hat{K}
+2 b_{k}^{2}\hat{U}_{k}\hat{K}\hat{U}_{k}-\Og b_{k}[\hat{U}_{k},\hat{K}]}
{\Og(4 b_{k}^{2}-\Og^{2})}\equiv \mathscr{L}[\hat{K}].
\label{eq:equation-g}
\end{equation}
An iteration procedure to solve Eq. (\ref{eq:equation-g}) has been
outlined in Ref.[\onlinecite{mishchenko-prl-2004}]. We adopt the
procedures here: according to the partition scheme of $\hat{K}$,
we use $\hat{K}^{(0)}$ to obtain the zeroth order approximation
given by $\hat{g}^{(0)}\equiv \msc{L}[\hat{K}^{(0)}(\hat{\rho})]$
which replaces $\hat{g}$ in $\hat{K}^{(1)}$ to generate a
correction due to the gradient term, \ie
$\hat{K}^{(1)}(\hat{g}^{(0)})$; we further insert
$\hat{K}^{(1)}(\hat{g}^{(0)})$ back to Eq. (\ref{eq:equation-g})
to obtain a correction $\msc{L}[\hat{K}^{(1)}(\hat{g}^{(0)})]$;
then we have the first order approximation to the quasiclassical
distribution function,
\begin{equation}
\hat{g}^{(1)}=\hat{g}^{(0)}+\msc{L}[\hat{K}^{(1)}(\hat{g}^{(0)})].
\end{equation}
The above procedure is repeated to a desired order using
\begin{equation}
\hat{g}^{(n)}=\hat{g}^{(n-1)}+\msc{L}[\hat{K}^{(1)}(\hat{g}^{(n-1)})].
\end{equation}
In this paper, the second order approximation is sufficient. The
full expression of the second order approximation for $\hat{g}$ is
tedious thus to be excluded in the following. The diffusion
equation is derived by an angle averaging in momentum space, which
allows all terms that are of odd order in $k_{i}$ ($i=x,y$) to
vanish while the combinations such as $k_{i}k_{j}$ contribute to
the averaging by a factor $k_{F}^{2}\dlt_{ij}$. \cite{rammer-book}

A Fourier transform from frequency domain back to the real time
brings us a diffusionlike equation for the density matrix,
\begin{eqnarray}
\frac{\pat}{\pat t}\hat{\rho}(t) & = &D\nabla^{2}\hat{\rho}
-\frac{1}{\tau_{\rm dp}}\hat{\rho}+\frac{1}{2\tau_{\rm dp}}
(\hat{\bm{z}}\times\hat{\bm{\sg}})\cdot\hat{\rho}(\hat{\bm{z}}\times\hat{\bm{\sg}})\nn\\
&& +iC
\left[\hat{\bm{z}}\times\hat{\bm{\sg}},\bm{\nabla}\hat{\rho}\right]
-B \left\{\hat{\bm{z}}\times\hat{\bm{\sg}},\bm{\nabla}\hat{\rho}\right\}\nn\\
&&+\Gamma \left[(\bm{m}\times\bm{\nabla})_{z}\hat{\rho}
-\hat{\bm{\sg}}\cdot\bm{m}\bm{\nabla}\hat{\rho}\cdot(\hat{\bm{z}}\times\hat{\bm{\sg}})
-(\hat{\bm{z}}\times\hat{\bm{\sg}})\cdot\bm{\nabla}\hat{\rho}\hat{\bm{\sg}}\cdot\bm{m}\right]\nn\\
&&+\frac{1}{2
\tau_{\vp}}\left(\hat{\bm{\sg}}\cdot\bm{m}\hat{\rho}\hat{\bm{\sg}}\cdot\bm{m}-\hat{\rho}\right)
-i\tilde{\Dlt}_{xc}[\hat{\bm{\sg}}\cdot\bm{m},\hat{\rho}]\nn\\
&&-2R
\left\{\hat{\bm{\sg}}\cdot\bm{m},(\bm{m}\times\bm{\nabla})_{z}\hat{\rho}\right\},
\label{eq:eom-leading-order}
\end{eqnarray}
where $\hat{\rho}$ assumes an energy dependence $\hat{\rho}\equiv
\hat{\rho}(\eps)$. The subscript is omitted for the brevity of
notation. In a two-dimensional system, the diffusion constant
$D=\tau v_{F}^{2}/2$ is given in terms of Fermi velocity $v_{F}$
and momentum relaxation time $\tau$. The renormalized exchange
splitting reads $\tilde{\Delta}_{xc}=(\Dlt_{xc}/2)/(4\vs^2+1)$,
where $\vs^{2}=(\Dlt_{xc}^{2}/4+\alp^{2}k_{F}^{2})\tau^{2}$. The
other parameters are given by
\begin{eqnarray}
C &=& \frac{\alp k_{F}v_{F}\tau}{(4\vs^{2}+1)^{2}},~
\Gamma=\frac{\alpha\Dlt_{xc}v_Fk_F\tau^2}{2(4\vs^2+1)^2}, R
=\frac{\alp\Dlt_{xc}^{2}\tau^{2}}{2(4\vs^{2}+1)},~
B =\frac{2\alpha^3 k_F^2\tau^2}{4\vs^2+1},\nn\\
\frac{1}{\tau_{\rm dp}} &=&
\frac{2\alp^{2}k_{F}^{2}\tau}{4\vs^{2}+1},~
\frac{1}{\tau_\varphi}=\frac{\Dlt_{xc}^2\tau}{4\vs^2+1}.\nn
\end{eqnarray}
$\tau_{\rm dp}$ is the relaxation time due to the D'yakonov-Perel
mechanism \cite{dp-she-1971} and $\tau_{\Dlt}\equiv 1/\Dlt_{xc}$
sets the time scale for the coherent precession of the spin
density around the magnetization. Equation
(\ref{eq:eom-leading-order}) is valid in the dirty limit $\vs\ll
1$, which enables the approximation $1+4\vs^{2}\approx 1$. The
charge density $n$ and spin density $\bm{S}$ are introduced by a
vector decomposition of the density matrix
$\hat{\rho}_{\eps}=n_{\eps}/2+\bm{S}_{\eps}\cdot{\hat{\bm\sg}}$.
In real experiments, \cite{mihai1,mihai2,miron-nature-2011} spin
transport in a ferromagnetic film experiences random magnetic
scatterers, for which we introduce phenomenologically an isotropic
spin-flip relaxation $\bm{S}/\tau_{sf}$.

After an integration over energy $\eps$, \ie $n=\mac{N}\int d\eps
~n_{\eps}$ and $\bm{S}=\mac{N}\int d\eps~\bm{S}_{\eps}$, we obtain
a set of diffusion equations for the charge and spin densities,
\ie
\begin{equation}
\frac{\pat n}{\pat t} = D\nabla^{2}n+ B {\bm\nabla}_z\cdot{\bm S}
+\Gamma {\bm\nabla}_z\cdot {\bm m} n
+R  {\bm\nabla}_z\cdot \bm{m}(\bm{S}\cdot \bm{m}),
\label{eq:charge-diffusion}
\end{equation}
where $\bm{\nabla}_{z}\equiv \hat{\bm{z}}\times\bm{\nabla}$ and
\begin{eqnarray}
\frac{\pat \bm {S}}{\pat t} &=& D \nabla^{2}{\bm S}
-\frac{\bm{S}}{\tau_{sf}}
-\frac{\bm{S}+S_z\bm{z}}{\tau_{\rm dp}} -\frac{1}{\tau_\Delta}\bm{S}\times \bm{m}
-\frac{\bm{m}\times(\bm{S}\times \bm{m})}{\tau_\varphi} \nn\\
&&+B {\bm\nabla}_z n +2 C {\bm\nabla}_z\times{\bm S}
+2 R ( \bm{m}\cdot{\bm\nabla}_z n) \bm{ m}\nn\\
&&+\Gamma \left[ \bm{m}\times({\bm\nabla}_z\times{\bm S})
+{\bm\nabla}_z\times({\bm m}\times{\bm S})\right].
\label{eq:spin-dynamics}
\end{eqnarray}
The anisotropy in spin relaxation is embedded naturally in our
model: the spin density components $S_{x}\hat{\bm{x}}$ and
$S_{y}\hat{\bm{y}}$ are relaxed at a rate $1/\tau_{\rm
dp}+1/\tau_{sf}$, while $S_{z}\hat{\bm{z}}$ is submitted to a
higher rate $2/\tau_{\rm dp}+1/\tau_{sf}$.

Equations (\ref{eq:charge-diffusion}) and (\ref{eq:spin-dynamics})
comprise one of the most important results in this paper. For a
broad range of the relative strength between the spin-orbit
coupling and exchange splitting $\alp k_{F}/\Delta_{xc}$, Eqs.
(\ref{eq:charge-diffusion}) and (\ref{eq:spin-dynamics}) not only
describe the spin dynamics in a ferromagnetic film but also
capture the symmetry of the spin-orbit torque. When the magnetism
vanishes $\Dlt_{xc}=0$, the $B$ term behaves as a source that
generates spin density electrically.
\cite{edelstein-1989,mishchenko-prl-2004} On the other hand, when
the Rashba spin-orbit coupling is absent ($\alpha=0$), the first
two lines in Eq. (\ref{eq:spin-dynamics}) describe a diffusive
motion of spin density in a ferromagnetic metal, which, to be
shown in the next section, agrees with early results.
\cite{zhang-li-2004} The $C$ term describes the coherent
precession of the spin density around the effective Rashba field.
The spin density induced by the Rashba field precesses around the
exchange field, which is described by the $\Gamma$ term, and is
thus at a higher order than the $C$ term in the dirty limit, for
$\Gamma=\Dlt_{xc}\tau C/2$. The $R$ term contributes to a
magnetization renormalization.

We shall assign a proper physical meaning to the \ti{transverse
spin dephasing time} $\tau_{\vp}$ defined in this paper. Here, the
dephasing time $\tau_{\vp}$ is different from the transverse spin
scattering time in, for example, Eq. (34) in
Ref.[\onlinecite{tserk-prb-2009}] that describes the disorder
contribution to the transverse spin scattering. $\tau_{\vp}$
rather contributes to the transverse spin conductivity
$\sg_{\tx{tr}}\propto \frac{n}{m}\tau_{\vp}$ and it plays the same
role to the transverse component of spin current as the momentum
relaxation time $\tau$ does in the ordinary Drude conductivity. In
fact, $\tau_{\vp}$ agrees with the calculation in
Ref.[\onlinecite{tserk-prb-2009}] when the weak ferromagnet limit
is taken, \ie $\mu_{\up}\approx \mu_{\dn}\approx\eps_{F}$ and
$\nu_{\up}\approx \nu_{\dn}\approx \mac{N}$.

\section{Spin Transport}
\subsection{Edelstein effect and spin-Hall effect: vanishing magnetism}
\label{sec:she} An electrically generated nonequilibrium spin
density due to spin-orbit coupling \cite{edelstein-1989} can be
extracted from Eq. (\ref{eq:spin-dynamics}) by setting the
exchange interaction to zero $\Dlt_{xc}=0$. If we keep
D'yakonov-Perel as the only spin relaxation mechanism and let
$\tau_{sf}=\infty$, Eq. (\ref{eq:spin-dynamics}) reads
\begin{equation}
D\bm{\nabla}^2\bm{S} - \frac{{\bm S}+S_z{\hat{\bm z}}}{\tau_{\rm
dp}} +2C \bm{\nabla}_{z}\times{\bm S} +B \bm{\nabla}_{z} n=0,
\label{eq:spin-hall-effect}
\end{equation}
which also describes the spin-Hall effect in the diffusive regime.
\cite{mishchenko-prl-2004,burkov-prb-2004,adagideli-bauer-prl-2005}
Besides the spin relaxation, the second term in Eq.
(\ref{eq:spin-hall-effect}), the spin dynamics is controlled by
two competing effects: the spin precession around the Rashba field
(third term) and the electrical spin generation first pointed out
by Edelstein.\cite{edelstein-1989} In an infinite medium where a
charge current is flowing along the $\hat{\bm{x}}$ direction, Eq.
(\ref{eq:spin-hall-effect}) leads to a solution
\begin{equation}
{\bm S}= e E \tau_{\rm dp} B \frac{n}{\eps_{F}} \hat{\bm{y}} =\frac{e
E\zeta}{\pi v_{F}}\hat{\bm{y}}, \label{eq:spin-density-pure-soi}
\end{equation}
where only the linear term in electric field has been retained. On
the right-hand side, we have used the charge density in a
two-dimensional system $n=k_{F}^{2}/(2\pi)$ and introduced the
parameter $\zeta=\alp k_{F}\tau$ as used in
Ref.[\onlinecite{mishchenko-prl-2004}]. In the presence of a weak
spin-orbit coupling, only the spin precession term survives; the
electrical spin generation dominates when the coupling is strong.

\subsection{Spin diffusion in a ferromagnet}
\label{sec:diff-ferro} Spin diffusion in a ferromagnet has been
discussed actively in the field of spintronics
\cite{zhang-li-2004,tserk-prb-2009,zhang-levy-fert-2002,tserk-rmp-2005}.
In this section we show explicitly that, by suppressing the Rashba
spin-orbit coupling, Eq. (\ref{eq:spin-dynamics}) is able to
describe spin diffusion in a ferromagnetic metal. A vanishing
Rashba spin-orbit coupling means $\alp=0$ and Eq.
(\ref{eq:spin-dynamics}) reduces to
\begin{equation}
\frac{\pat\bm{S}}{\pat t} = D\nabla^{2}\bm{S}
+\frac{\bm{m}\times\bm{S}}{\tau_{\Dlt}}
-\frac{\bm{S}}{\tau_{sf}}
- \frac{{\bm m}\times({\bm S}\times{\bm m})}{\tau_\varphi},
\end{equation}
This equation only differs from the result of Ref.
[\onlinecite{zhang-levy-fert-2002}] by a dephasing term of the
transverse component of the spin density.

In a ferromagnetic metal, we may divide the spin density into a
\ti{longitudinal} component that follows the magnetization
direction adiabatically, and a deviation that is
\ti{perpendicular} to the magnetization, \ie
$\bm{S}=s_{0}\bm{m}+\dlt\bm{S}$ where $s_{0}$ is the local
equilibrium spin density. Such a partition, after restoring the
electric field by $\bm{\nabla}\raw\bm{\nabla}+(e/\eps_{F})\bm{E}$,
gives rise to
\begin{eqnarray}
\frac{\pat}{\pat t}\dlt\bm{S}+\frac{\pat}{\pat t}s_{0}\bm{m}
&=&s_{0}D\nabla^{2}\bm{m}+D\nabla^{2}\dlt\bm{S}+D e P_{F}\mac{N}_{F}\bm{E}\cdot\bm{\nabla}\bm{m}\nn\\
&&-\frac{\dlt\bm{S}}{\tau_{sf}}-\frac{s_{0}\bm{m}}{\tau_{sf}}-\frac{\dlt\bm{S}}{\tau_\varphi}
+\frac{1}{\tau_\Delta}\bm{m}\times\dlt\bm{S},
\label{eq:full-spinaccu-dynamics}
\end{eqnarray}
where the magnetic order parameter is allowed to be spatial
dependent $\bm{m}=\bm{m}({\bm r},t)$. We introduce $P_{F}$ the
spin polarization and $\mac{N}_{F}$ the density of state; both are
at Fermi energy $\eps_{F}$.

In a smooth magnetic texture, the characteristic length scale of
the magnetic profile is much larger than the length scale of
electron transport; we discard the contribution
$D\nabla^{2}\dlt\bm{S}$.\cite{zhang-li-2004} The diffusion of the
equilibrium spin density follows $s_{0}D\nabla^{2}\bm{m}\approx
s_{0}\bm{m}/\tau_{sf}$. In this paper, we retain only terms that
are at first order in temporal derivative, which simplifies Eq.
(\ref{eq:full-spinaccu-dynamics}) to
\begin{equation}
\label{eq:zl1}
\xi\frac{\dlt\bm{S}}{\tau_{\Dlt}}
-\frac{\bm{m}\times\dlt\bm{S}}{\tau_{\Dlt}}=
D e P_{F}\mac{N}_{F}\bm{E}\cdot\bm{\nabla}\bm{m}
-s_{0}\frac{\pat\bm{m}}{\pat t}.
\end{equation}
The last equation can be solved exactly to show
\begin{eqnarray}
\label{eq:zl2-spin-accu}
\dlt\bm{S} = \frac{\tau_{\Dlt}}{1+\xi^2} && \left[\frac{P_F}{e}{\bm
m}\times({\bm j}_e\cdot{\bm\nabla}){\bm m} +\xi
\frac{P_F}{e}({\bm j}_e\cdot{\bm\nabla}){\bm m}\right.\nn\\&&\left.-s_{0}{\bm
m}\times\frac{\pat\bm{m}}{\pat t}-\xi
s_{0}\frac{\pat\bm{m}}{\pat t}\right]
\end{eqnarray}
where $\xi=\tau_\Dlt(1/\tau_{sf}+1/\tau_\varphi)$ and the electric
current $\bm{j}_{e}=e^{2}n\tau\bm{E}/m$ is given in terms of
electron density $n$. Apart from the transverse dephasing time
absorbed in parameter $\xi$, the nonequilibrium spin density Eq.
(\ref{eq:zl2-spin-accu}) agrees with Eq. (8) in
Ref.[\onlinecite{zhang-li-2004}]. Given the knowledge of the
nonequilibrium spin density, the spin torque, defined as
\begin{equation}
{\bm T}=-\frac{1}{\tau_{\Dlt}}\bm{m}\times\dlt{\bm S}+\frac{1}{\tau_\varphi}\dlt{\bm S},
\label{eq:spin-torque-definition}
\end{equation}
is given by
\begin{eqnarray}
\label{eq:zl2-torque} {\bm T}=&& -(1-\xi\tilde{\beta})
s_{0}\frac{\pat\bm{m}}{\pat t} +\tilde{\beta} s_{0}{\bm
m}\times\frac{\pat\bm{m}}{\pat t} \nn\\
&& +(1-\xi\tilde{\beta}) \frac{P_F}{e}({\bm
j}_e\cdot{\bm\nabla})\bm{m} -\tilde{\beta}
\frac{P_F}{e}\bm{m}\times({\bm j}_e\cdot{\bm\nabla})\bm{m}
\end{eqnarray}
where $\beta=\tau_{\Dlt}/\tau_{sf}$ and
$\tilde{\beta}=\beta/(1+\xi^2)$. By assuming a long dephasing time
of the transverse component $\tau_\varphi\raw\infty$ then
$\xi\approx \beta$, Eq.  (\ref{eq:zl2-torque}) reproduces Eq. (9)
in Ref.[\onlinecite{zhang-li-2004}]. On the other hand, a short
spin dephasing time $\tau_{\vp}\raw 0$  yields $\tilde{\beta}\raw
0$ which results in a pure adiabatic torque, \ie the torque
reduces to the first and third terms in Eq. (\ref{eq:zl2-torque}).

\section{Rashba spin torque}
\label{sec:rashba-spin-torque} The primary focus of this article
is the Rashba torque originating from the coexistence of magnetism
and Rashba spin-orbit coupling. In this section, we apply Eqs.
(\ref{eq:charge-diffusion}) and (\ref{eq:spin-dynamics}) to study
the properties of this torque and concentrate on possible
analytical aspects in the bulk system or an infinite medium.
Analytical results provide a better understanding of the physical
processes behind the Rashba torque and a more transparent view on
the structure of the diffusion equations derived in Sec.
\ref{sec:diffusion-equation}. To serve this purpose, we first
derive a formula that characterizes the general symmetry and
angular dependence of the Rashba torque. Then for two limiting
cases at weak and strong spin-orbit couplings, we are able to
directly compare our results to experiments.

\subsection{General symmetry and angular dependence}
\label{sec:general-symmetry} Recent studies showed that the
spin-orbit torque in a ferromagnetic heterostructure possesses
peculiar symmetries with respect to magnetization inversion and a
complex angular dependence.\cite{garello} To be more specific, the
angular dependence discussed here refers to the experimental
observation that the torque amplitudes vary as functions of
magnetization direction. We demonstrate in the following that such
symmetries and angular dependence are encoded coherently in our
simple model.

Needless to say, finding a general analytical solution to Eq.
(\ref{eq:spin-dynamics}) with boundary conditions is by no means
an easy task. But, such solutions to the spin density and spin
torque do exist in an infinite medium and the behavior featured by
these solutions, as the numerical solutions suggest, persists into
a finite system.\cite{ortiz-apl-2013} We reorganize Eq.
(\ref{eq:spin-dynamics}) as
\begin{eqnarray}
&&\frac{\bm{S}}{\tau_{sf}} +\frac{\bm{S}+S_z
\hat{\bm{z}}}{\tau_{\rm dp}} +\frac{1}{\tau_\Delta}\bm{S}\times
\bm{m} +\frac{1}{\tau_\varphi}\bm{m}\times(\bm{S}\times
\bm{m})={\bm X},\label{eq:X}
\end{eqnarray}
where the right-hand side combines the time and spatial
derivatives of the spin and charge densities
\begin{eqnarray}
\bm{X} &\equiv & -\frac{\pat \bm {S}}{\pat t}+ D \nabla^{2}{\bm S}
+B {\bm\nabla}_z n +2 C {\bm\nabla}_z\times{\bm S}
+2 R ( \bm{m}\cdot{\bm\nabla}_z n) \bm{ m}\nn\\
&&+\Gamma \left[ \bm{m}\times({\bm\nabla}_z\times{\bm S})
+{\bm\nabla}_z\times({\bm m}\times{\bm S})\right].
\end{eqnarray}

A stationary state solution defined by $\pat\bm{S}/\pat t=0$ is of
our current interest, whereas, in the next section, we will see
that this term induces a correction contributing to the spin and
charge pumping effects. In an infinite medium with an applied
electric field $\bm{E}$, we again replace the spatial gradient
${\bm\nabla}$ by $(e/\eps_{F})\bm{E}$ and ${\bm X}$ reduces to
\begin{eqnarray}
\bm{X} \approx \frac{e}{\eps_{F}}
&& \left[B n \hat{\bm{z}}\times\bm{E}+2C(\hat{\bm{z}}\times\bm{E})\times\bm{S}
+2 R n \left(\bm{m}\cdot(\hat{\bm{z}}\times\bm{E})\right)\bm{m}\right.\nn\\
&&\left. +\Gm \bm{m}\times((\hat{\bm{z}}\times\bm{E})\times\bm{S})
+\Gm (\hat{\bm{z}}\times\bm{E})\times(\bm{m}\times\bm{S})\right]
\label{eq:xrt}
\end{eqnarray}
where we have discarded $\nabla^{2}\bm{S}$ that is quadratic in
$E$.

For a general expression Eq. (\ref{eq:xrt}), we may solve Eq.
(\ref{eq:X}) using the partition ${\bm S}=S_{\para}{\bm
m}+\dlt\bm{S}$. A lengthy algebra leads to
\begin{eqnarray}
S_\para &=& \tau_{\rm dp}\gamma_\theta\chi_\theta \bm{X}\cdot
\left[{\bm m}(1+\xi\chi\tilde{\beta}\sin^2\theta)\right.\nn\\
&&\left.-\chi\til{\beta}m_z(\xi{\bm m}\times\hat{\bm{z}}\times{\bm
m}+\hat{\bm{z}}\times{\bm m})\right],
\label{eq:spin-density-infinite-para}\\
\dlt\bm{S} &=& \frac{\tau_\Delta\gamma_\theta}{1+\xi^2} \left[{\bm
m}\times{\bm X}+\xi{\bm m}\times{\bm X}\times{\bm m}
-\xi\chi_\theta m_z({\bm X}\cdot{\bm m}){\bm m}\times\hat{\bm{z}}\times{\bm m}\right.\nn\\
&&\left.+\chi_\theta(m_z{\bm X}\cdot{\bm m}+\beta{\bm
X}\cdot(\hat{\bm{z}}\times{\bm m}))\hat{\bm{z}}\times{\bm
m}\right] \label{eq:spin-density-infinite-perp}
\end{eqnarray}
where $\theta$ is defined as the azimuthal angle between $\bm{m}$
and $\hat{\bm{z}}$ and
\begin{eqnarray}
&&\chi=\frac{\tau_{sf}}{\tau_{sf}+\tau_{\rm
dp}},\;\chi_\theta=\frac{\chi}{1+\chi\cos^2\theta},\;
\gamma_\theta=\frac{1}{1+\xi\tilde{\beta}\chi_\theta\sin^2\theta}.
\end{eqnarray}

The results of spin density in Eq.
(\ref{eq:spin-density-infinite-perp}) give rise to the spin torque
defined by Eq. (\ref{eq:spin-torque-definition}). In an infinite
medium, the spin torque reads
\begin{eqnarray}
{\bm T}&=&\gm_\theta\left[(1-\xi\tilde{\beta}){\bm m}\times{\bm
X}\times{\bm m}
+\tilde{\beta}{\bm m}\times{\bm X}\right.\nn\\
&&\left. +\til{\beta}\chi_\theta
[(\xi-\beta){\bm X}\cdot(\hat{\bm{z}}\times{\bm m})-m_z({\bm X}\cdot{\bm m})]
\hat{\bm{z}}\times{\bm m}\right.\nn\\
&&\left. -\chi_\theta[(1-\xi\tilde{\beta})m_z{\bm X}\cdot{\bm
m}+\tilde{\beta}{\bm X}\cdot(\hat{\bm{z}}\times{\bm m})]{\bm
m}\times\hat{\bm{z}}\times{\bm m}\right],\label{eq:tt}
\end{eqnarray}
which clearly exhibits two outstanding features. First, it is
possible to divide the torque into two components that are either
odd or even with respect to inversion of magnetization direction.
Second, every component has a pronounced angular dependence.

The formulation of Eq. (\ref{eq:tt}) motivates us to attempt an
interpretation of $\bm{X}$ as a \ti{source} term, which allows us
to extend the applicability of Eq. (\ref{eq:tt}) to include other
driving mechanisms due to temporal variation or magnetic texture.
We may further simplify Eq. (\ref{eq:xrt}) by observing that all
nonequilibrium spin and charge densities shall--in the lowest
order--be linear in $E$. As we are only interested in the linear
response regime, we can approximate $\bm{S}$ by an equilibrium
value $n P_{F}\bm{m}$ and the source term becomes
\begin{eqnarray}
\bm{X} \approx e \mac{N}_{F}&& \left[B \hat{\bm{z}}\times\bm{E}+2C
P_{F}(\hat{\bm{z}}\times\bm{E})\times\bm{m}
+2 R \left(\bm{m}\cdot(\hat{\bm{z}}\times\bm{E})\right)\bm{m}\right.\nn\\
&&\left. +\Gm P_{F}
\bm{m}\times((\hat{\bm{z}}\times\bm{E})\times\bm{m}) \right],
\label{eq:x-general-final}
\end{eqnarray}
which serves as a starting point of the following discussions on
spin torques in two major limits.

\subsection{Weak spin-orbit coupling}
\label{sec:weak-soi} In our system, a weak Rashba spin-orbit
coupling implies a low D'yakonov-Perel relaxation rate
$1/\tau_{\rm dp}\propto \alp^{2}$ such that $\tau_{\rm dp}\gg
\tau_{sf}, \tau_{\Dlt}$, indicating the spin relaxation is
dominated by random magnetic impurities. In this regime, spin
precession about the total field dominates the electrical spin
generation; we may retain only $C$ and $\Gamma$ terms in Eq.
(\ref{eq:x-general-final}) and discard $B$ and $R$ terms that are
of higher order in $\alp$. Therefore, when an electric field is
applied along the $\hat{x}$ direction, Eq.
(\ref{eq:x-general-final}) becomes
\begin{eqnarray}
\bm{X}\approx e \mac{N}_F P_F E\left[2C \hat{\bm{y}}\times{\bm
m}+\Gamma{\bm m}\times(\hat{\bm{y}}\times{\bm m})\right]
\end{eqnarray}
and the torque given in Eq. (\ref{eq:tt}) reduces to a commonly
accepted form
\begin{eqnarray}
{\bm T}=T_{\perp}\hat{\bm{y}}\times\bm{m}
+T_{\para}\bm{m}\times(\hat{\bm{y}}\times\bm{m}),
\label{torque:wr}
\end{eqnarray}
consisting of both out-of-plane ($T_\perp$) and in-plane
($T_\para$) components with magnitudes determined by
\begin{eqnarray}
T_\perp &=& e E P_{F}\mac{N}_F\left[2(1-\xi\tilde{\beta})
C+\tilde{\beta}\Gamma\right], \\
T_\para &=& e E P_{F}\mac{N}_F\left[(1-\xi\tilde{\beta})\Gamma - 2
\til{\beta} C\right]. \label{eq:torque-magnitude-weak-alp}
\end{eqnarray}
Note that the in-plane torque in Eq.
(\ref{eq:torque-magnitude-weak-alp}) may experience a sign flip,
depending on the competition between spin relaxation and
precession.

To compare directly with the results in
Ref.[\onlinecite{manchon-prb}], we allow $\tau_{sf}\raw \infty$
and $\tau_{\varphi}\raw \infty$, then $\beta\approx 0$. Under
these assumptions, we have $T_\perp\approx 2 e E P_{F}\mac{N}_F C$
and $T_\para\approx e E P_{F}\mac{N}_F\Gamma$. In the dirty limit,
$\Gamma\ll C$ due to $\Dlt_{xc}\tau\ll 1$. By making use of the
relation for the polarization $P_{F} = \Dlt_{xc}/\eps_{F}$ and the
Drude relation $j_{e}=e^{2}n\tau E/m$, we obtain the out-of-plane
torque
\begin{eqnarray}
\bm{T} =2\frac{\alp
m\Dlt_{xc}}{e\eps_{F}}j_{e}\hat{\bm{y}}\times\bm{m},
\end{eqnarray}
which agrees with the spin torque in an infinite system in the
corresponding limit as derived in Ref.[\onlinecite{manchon-prb}].

\subsection{Strong spin-orbit coupling}
\label{sec:strong-soi} In the presence of a strong spin-orbit
coupling, two effects are dominating: electric generation of spin
density \cite{edelstein-1989} and D'yakonov-Perel spin relaxation
mechanism.\cite{dp} As the electric field is aligned along the
$\hat{\bm{x}}$ direction, $\bm{X}$ is simplified to be
\begin{eqnarray}
\bm{X} \approx e \mac{N}_{F} E && \left[B \hat{\bm{y}}+2C
P_{F}\hat{\bm{y}}\times\bm{m} +2 R m_{y}\bm{m} +\Gm P_{F}
\bm{m}\times(\hat{\bm{y}}\times\bm{m}) \right],
\end{eqnarray}
and the corresponding spin torque is
\begin{eqnarray}
\bm{T}
=&& \gamma_\theta(T_{\perp}^0\hat{\bm{y}}\times\bm{m}+T_{\para}^0\bm{m}\times\hat{\bm{y}}\times\bm{m})\nn\\
&&+\gamma_\theta\chi_\theta (T_{\perp}^x m_{x} + T_{\perp}^{yz}m_{y} m_{z})\hat{\bm{z}}\times\bm{m}\nn\\\
&&+\gamma_\theta\chi_\theta (T_{\para}^x m_{x} +
T_{\para}^{yz}m_{y} m_{z})\bm{m}\times(\hat{\bm{z}}\times\bm{m}),
\label{torque:general}
\end{eqnarray}
where the torque amplitude parameters are defined as
\begin{eqnarray}
T_\para^0 &=& eE{\cal N}_F[2(1-\xi\tilde{\beta}) C P_F+\tilde{\beta}(B+\Gamma P_F)],\nn\\
T_\perp^0 &=& eE{\cal N}_F[-2\tilde{\beta} C P_F+(1-\xi\tilde{\beta})(B+\Gamma P_F)],\nn\\
T_\para^x &=& -eE{\cal N}_F \tilde{\beta}(B+\Gamma P_F),\nn\\
T_\para^{yz} &=& eE{\cal N}_F [2\tilde{\beta}CP_F-(1-\xi\tilde{\beta})(B+2R)],\nn\\
T_\perp^x &=& -eE{\cal N}_F\tilde{\beta}(\xi-\beta)(B+\Gamma P_F),\nn\\
T_\perp^{yz} &=& -eE{\cal N}_F
[2(\xi-\beta)CP_F+(1-\xi\tilde{\beta})(B+2R)].
\end{eqnarray}

Equation (\ref{torque:general}) comprises another major result of
this paper. The first term, an out-of-plane torque, can be
understood as a fieldlike torque produced through the ferromagnet
and the spin density generated by the inverse spin-galvanic
effect. The second term, an in-plane torque, originates from the
Slonczewski-Berger-type spin-transfer torque that requires spin
dephasing of the transverse component of spin density. The last
two terms $\hat{\bm{z}}\times\bm{m}$ and
$\bm{m}\times(\hat{\bm{z}}\times\bm{m})$ are governed by the
anisotropy in spin relaxation which allows the generation of spin
density components to be perpendicular to both $\bm{m}$ and the
effective Rashba field. In general, the relative magnitude of
these different terms are material dependent.

In fact, the symmetry reflected in Eq. (\ref{torque:general})
compares favorably to the spin torque formula proposed based on
experiments by Garello \ti{et al}. \cite{garello} More
interestingly, if we allow the anisotropy in spin relaxation time
to vanish by taking $\tau_{\rm dp}\gg\tau_{sf}$, Eq.
(\ref{torque:general}) reduces to the form of Eq.
(\ref{torque:wr}) consisting only of the in-plane and out-of-plane
components, whereas the complex angular dependence diminishes
accordingly. This is a strong indication that this angular
dependence discovered in our model arises from the anisotropic
spin relaxation. Meanwhile, such an angular dependence obtained
here in an infinite medium persists into a realistic experimental
setup with boundaries and it is insensitive to the change in
sample size.\cite{ortiz-apl-2013}

\section{Numerical results} \label{sec:numeric}
In previous sections, analytical results for the spin density and
Rashba torque were obtained in various limits with respect to the
relative magnitude between the spin-orbit coupling and exchange
field. In this section, we numerically solve Eqs.
(\ref{eq:charge-diffusion}) and (\ref{eq:spin-dynamics}) to
demonstrate that they provide a coherent framework to describe the
spin dynamics as well as spin torques in the diffusive regime for
a wider range of parameters. Here, we consider an in-plane
magnetization that lies along the $\hat{\bm{x}}$ direction and
another case where the magnetization is perpendicular to the thin
film plane is reported elsewhere.\cite{wang-manchon-2011} For such
a two-dimensional electron system, we adopt the following boundary
conditions. First, we enforce a vanishing spin accumulation at the
edges along the transverse direction, \ie $S(y=0,L)=0$. This
condition implies a strong spin-flip scattering at the edges,
which is consistent with the experimental observations in
spin-Hall effect.\cite{kato-science-2004} Second, an electric
field is applied along the $\hat{\bm{x}}$ direction; therefore, we
set the charge densities at two ends of the propagation direction
to be constant $n_L=n_R=n_F$. The second boundary condition sets
the charge density at the Fermi level. Equivalently, one can apply
a voltage drop along the transport direction instead of an
explicit inclusion of an electric field.

\begin{figure}
\centering
\includegraphics[trim = 0mm 0mm 0mm 0mm, clip, scale=0.5]{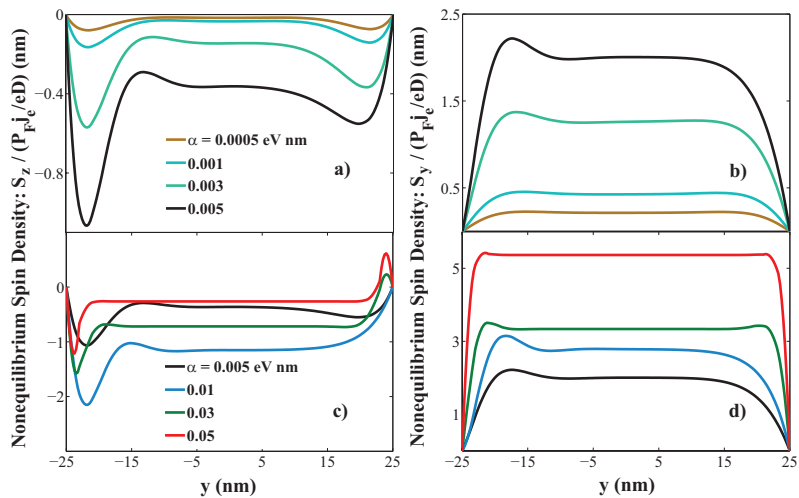}
\caption{\label{fig:spins} (Color online) Spatial profile of the
nonequilibrium spin density $S_z$ (a),(c) and $S_{y}$ (b),(d) for
various values of the Rashba constant. The width of the wire is
$L=50$~nm. The magnetization direction is along the $\hat{\bm{x}}$
axis. Other parameters are momentum relaxation time
$\tau=10^{-15}$ s, exchange splitting $\tau_\Delta=10^{-14}$ s,
spin relaxation time $\tau_{sf}=10^{-12}$ s, and the Fermi vector
$k_{F}=4.3~\tx{nm}^{-1}$.}
\end{figure}

The numerical results of the spin densities are summarized in Fig.
\ref{fig:spins}. From the top panels [(a),(b)] to the lower ones
[(c),(d)], for a fixed exchange splitting, the system transitions
from a weak (spin-orbit-) coupling regime to a strong coupling
regime. To illustrate this transition, the $S_z$ component of the
spin density evolves from a symmetric spatial distribution in the
weak spin-orbit-coupling regime, with $\alpha=5\times
10^{-4}~\rm{eV}~\rm{nm}$ in Fig. 1(a), to an antisymmetric spatial
distribution in the strong coupling regime, with $\alpha=5\times
10^{-2}~\rm{eV}~\rm{nm}$ in Fig. 1(c). Note that throughout this
transition, the in-plane spin density $S_y$ is robust yet roughly
constant in the bulk.

This change in symmetry and the emergence of peaks close to the
boundaries are resulting from the competition between the Rashba
and exchange fields. In the weak coupling regime, the total field
is dominated by the exchange field pointing at the $\hat{\bm{x}}$
direction, about which the spin density profile is symmetric in
space. As the spin-orbit coupling increases, the total field is
tilted towards the $\hat{\bm{y}}$ axis; then the spin projections
along $+y$ and $-y$ are no longer symmetric, as indicted by curves
with intermediate $\alp$ values in Figs. \ref{fig:spins}(a) and
\ref{fig:spins}(b). In the strong coupling regime, when the Rashba
coupling overrules the exchange field, the antisymmetric profile
of $S_{z}$ and the symmetric one of $S_{y}$ follow naturally from
the spin-Hall effect induced by the spin-orbit interaction.

\begin{figure}
\centering
\includegraphics[trim = 0mm 0mm 0mm 0mm, clip, scale=0.5]{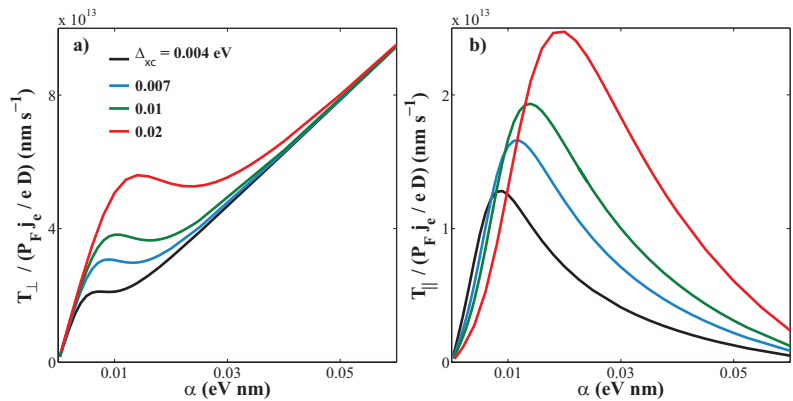}
\caption{\label{fig:torques} (Color online) Magnitude of the
out-of-plane torque $T_{\perp}$ (a) and in-plane torque
$T_{\para}$ (b) as a function of Rashba constant for various
exchange splitting. Other parameters are the same as in Fig.
\ref{fig:spins}.}
\end{figure}

The out-of-plane and in-plane torques are plotted in Fig.
\ref{fig:torques} with respect to the Rashba constant $\alp$ for
various exchange splittings. The transition regions are of
particular interest. During the transition from a weak to strong
coupling, see Fig. \ref{fig:torques}(a), the magnitude of the
out-of-plane torque $T_{\perp}$ first reaches a plateau, then
rises again as $\alp$ increases. In the large $\alp$ limit, though
the magnitude of the torque increases with $\alp$, the torque
efficiency defined as $dT_{\perp}/d\alp$ is actually smaller than
it is in the weak coupling. This picture is consistent with the
semiclassical Boltzmann equation description in Ref.
[\onlinecite{manchon-prb}]. This behavior is caused by the
different processes generating the Rashba torque in both regimes.
As discussed in Sec. \ref{sec:weak-soi} and Sec.
\ref{sec:strong-soi}, in the weak coupling regime, the torque is
dominated by the spin precession around the Rashba field, whereas
in the strong coupling, the electrical generation of spin density
dominates. These two distinct processes show different
efficiencies.

The in-plane torque $T_{\para}$ behaves differently. In the strong
coupling limit, $T_{\para}$ is proportional to $1/\alp$ due to the
large D'yakonov-Perel spin relaxation rate that is of order
$\alp^{2}$. A stronger spin-orbit coupling therefore means a
decrease in the torque magnitude. The transition suggests that the
optimal magnitude of the in-plane torque is achieved when the
exchange energy is about the same order of magnitude as the Rashba
splitting $\alp k_{F}$.

\section{Dynamics}
\label{sec:dynamics} Our focus in the previous sections has always
been on a stationary state with a homogeneous magnetization and
the temporal and spatial variations of the ferromagnetic order
parameter are neglected entirely. In this section, we demonstrate
that the formulation outlined in Eq. (\ref{eq:spin-dynamics}) is
able to address dynamic effects such as spin pumping and
magnetization damping. We shall consider only the adiabatic limit
where the frequency of magnetization motion is much lower than
that of any electronic processes. Without the loss of generality,
the anisotropy in spin relaxation is suppressed, for we are keen
to provide a qualitative picture rather than pinpointing
subtleties.

\subsection{Spin and charge pumping}
Now we consider a homogeneous single-domain ferromagnet with a
moving magnetization in the absence of external electric field. In
the adiabatic limit, while treating the spin-orbit coupling as a
perturbation, the lowest-order correction to the spin dynamics is
to let the source term
\begin{equation}
\bm{X} \approx -\pat_{t}{\bm S}.
\end{equation}
The magnetization motion brings the system out of equilibrium and
induces a nonequilibrium spin density. We can no longer naively
assume that the spin density is always following the magnetization
direction. To get the nonequilibrium part, we perform the usual
decomposition $\bm{S} = s_{0}\bm{m}+\dlt\bm{S}$ as in Sec.
\ref{sec:diff-ferro}, where $\dlt\bm{S}$ is referring to the
nonequilibrium part induced by the magnetization motion. Here, we
neglect terms like $D\nabla^{2}\dlt\bm{S}$ and a simple algebra
leads to
\begin{equation}
\dlt\bm{S} = \frac{\tau_{\Dlt} s_{0}}{1+\xi^{2}} \left({\bm
m}\times\frac{d\bm{m}}{dt}+\xi\frac{d \bm{m}}{dt}\right).
\label{eq:spin-pumping-spin-density}
\end{equation}

Equation (\ref{eq:spin-pumping-spin-density}) is a formal analogy
to the conventional spin pumping theory developed in magnetic
multilayers using the scattering matrix approach.
\cite{tserk-rmp-2005, tserk-spin-pumping-prl-2002} Two components
exist in the pumping-induced spin density and both of them are
perpendicular to the magnetization direction. In the absence of
spin-flip scattering $\tau_{sf}\raw \infty$ thus $\xi\ll 1$ (in
the dirty limit considered here), the first term
$\bm{m}\times\dot{\bm{m}}$ dominates. In the conventional spin
pumping theory, this contribution is governed by the real part of
the spin-mixing conductance that is usually much larger than its
imaginary counterpart associated with $\dot{\bm{m}}$. Equation
(\ref{eq:spin-pumping-spin-density}) seems to suggest a similar
trend. On the other hand, a strong spin-flip scattering is
expected to be detrimental to the nonequilibrium spin density,
which is also encoded in Eq. (\ref{eq:spin-pumping-spin-density}):
the magnitude of $\dlt\bm{S}$ decreases when the spin-flip
relaxation rate $1/\tau_{sf}$ increases.

Furthermore, the spin density induced by the magnetization motion
generates a charge current via the spin galvanic effect,
\cite{Ganichev} which can be estimated qualitatively to be
\begin{equation}
{\bm J}_c \propto \frac{\alp \tau_\Dlt s_{0}}{1+\xi^{2}} \hat{{\bm
z}}\times\left({\bm m}\times\frac{d\bm{m}}{dt}+\xi\frac{d
\bm{m}}{dt}\right)
\end{equation}
The magnitude of the charge current is proportional to the
frequency of the magnetization precession.

\subsection{Magnetic damping}
When a magnetization moves in a sea of itinerant electrons, the
coupling between the localized and itinerant electrons induces a
\ti{friction} to this motion. This friction has been described in
terms of the reciprocal of the spin pumping in a magnetic texture.
\cite{zz2009} The dynamical motion of the magnetic texture pumps a
spin current that contributes to a magnetic damping when
reabsorbed by the texture. In the present case, we show that the
pumping of a spin-polarized current studied above can also
contribute to the magnetic damping following the same process. In
order to describe a magnetic texture, we allow the magnetization
direction to assume a spatial dependence, i.e.,
$\bm{m}=\bm{m}(\bm{r})$. We limit ourselves to a weak Rashba
spin-orbit coupling in order to avoid the complexity due to
anisotropy in spin relaxation. For the present purpose, we keep
the following \ti{source} term:
\begin{equation}
\bm{X} = -\partial_t{\bm S}+D{\bm \nabla}^2{\bm S}+2C({\bm
z}\times{\bm \nabla})\times{\bm S}. \label{eq:x-damping}
\end{equation}

To be more specific, we identify two sources for spatially
dependent magnetic damping. One comes from the interplay between
the diffusive spin dynamics and the magnetization motion, i.e.,
the second term in Eq. (\ref{eq:x-damping}). The other apparently
attributes to the Rashba torque, i.e., the third term in Eq.
(\ref{eq:x-damping}). We consider here an adiabatic magnetization
dynamics, meaning that the electronic spin process, characterized
by a time scale $\tau_\Delta$, is the fastest whereas the
magnetization motion, with a time scale $\tau_M$, is the slowest.
Without loss of generality, we allow the spin dephasing time to
sit in between, i.e., $\tau_\Delta\ll\tau_{\varphi}\ll\tau_M$.
Under these assumptions, the nonequilibrium spin density pumped by
the magnetization motion reads $\delta {\bm S}\approx
-s_0\tau_\Delta{\bm m}\times\partial_t{\bm m}$ and the spatial
dependent damping torques are given by
\begin{eqnarray}
{\bm T}_D &=&-s_0\tau_\Delta D{\bm m}\times[{\bm m}\times{\bm \nabla}^2({\bm m}\times\partial_t{\bm m})],
\label{eq:diffusion-damping-torque}\\
{\bm T}_R &=&2 s_0\tau_\Delta C{\bm m}\times\{{\bm
m}\times\left[({\bm z}\times{\bm \nabla})\times({\bm
m}\times\partial_t{\bm
m})\right]\}\label{eq:rashba-damping-toque}.
\end{eqnarray}
It is worth pointing out the symmetry properties of the last two
damping torques. The damping torque due to spin diffusion, ${\bm
T}_D$, is second order in the spatial gradient and is thus
invariant under spatial inversion ${\bm \nabla}\rightarrow-{\bm
\nabla}$. In fact, Eq.(\ref{eq:diffusion-damping-torque}) has the
same symmetry as the damping torque obtained by Zhang and
Zhang.\cite{zz2009} The other damping torque that arises from the
Rashba spin-orbit coupling, ${\bm T}_R$, is only \ti{linear} in
spatial gradient and is therefore referred to as {\em chiral}. In
other words, in contrast to ${\bm T}_D$, the magnetic damping due
to Rashba spin-orbit coupling is antisymmetric upon spatial
inversion. Equation (\ref{eq:rashba-damping-toque}) is in
agreement with the damping formula derived by Kim \ti{et
al.}\cite{kim-prl-2012} Moreover, a more complex angular
dependence of the damping coefficient emerges when the
D'yakonov-Perel anisotropic spin relaxation is taken into account.

\section{Discussion}
\label{sec:discussion} Current-induced magnetization dynamics in a
single ferromagnetic layer has been observed in various structures
that involve interfaces between transition metal ferromagnets,
heavy metals, and/or metal-oxide insulators. Existing experimental
systems are
Pt/Co/AlO$_x$,\cite{mihai1,mihai2,miron-nature-2011,pi}
Ta/CoFeB/MgO,\cite{suzuki} and Pt/NiFe and Pt/Co
bilayers,\cite{liu} as well as dilute magnetic semiconductors such
as (Ga,Mn)As. \cite{chernyshov,fang} Besides the structural
complexity in such systems, an unclear picture of spin-orbit
coupling in the bulk as well as at interfaces places a challenge
to unravel the nature of spin-orbit torque.

\subsection{Validity of Rashba model in realistic interfaces}
The well-known Rashba-type effective interfacial spin-orbit
Hamiltonian was pioneered by E. I. Rashba to model the influence
of asymmetric interfaces in semiconducting two-dimensional
electron gas: \cite{rashba-soi} a sharp potential drop, emerging
at the interface (say, in the $x$-$y$ plane) between two
materials, gives rise to a potential gradient ${\bm\nabla}V$ that
is normal to the interface, \ie $\bm{\nabla}V\approx \xi_{\rm
so}(\bm{r})\hat{\bm{z}}$. In case a rotational symmetry exists
in-plane, a spherical Fermi surface assumption allows the
spin-orbit interaction Hamiltonian to have the form ${\hat
H}_{R}=\alp \hat{\bm{\sg}}\cdot(\bm{p}\times{\hat{\bm z}})$, where
$\alp \approx\langle\xi\rangle/4m^2c^2$. As a matter of fact, in
semiconducting interfaces where the transport is described by a
limited number of bands around a high symmetry point, the Rashba
form can be recovered by $\bm{k}\cdot\bm{p}$ theory.
\cite{winkler}

However, one can properly question the validity of the simple
Rashba spin-orbit-coupling model for interfaces involving heavy
metals and ferromagnets, where the band structure and Fermi
surfaces are much more complex than low doped semiconducting
two-dimensional electron gases for which it was initially
proposed. Nonetheless, the existence of a symmetry
breaking-induced spin splitting of the Rashba type has been well
established by angle-resolved photoemission spectroscopy in
systems consisting of a wide variety of metallic surfaces,
\cite{metals,Gd,Bi} quantum wells,\cite{qw} and even oxide
heterointerfaces.\cite{oxides} Several important works published
in the past few years on spin spirals induced by
Dzyaloshinskii-Moriya interaction at W/Fe and W/Mn interfaces
\cite{wies} also argue in favor of the presence of a sizable
Rashba-type spin-orbit coupling.

Besides the aforementioned experimental investigations performed
on clean and epitaxially grown systems, efforts in numerical
calculations have been made to the identification of an asymmetric
spin splitting in the band structure of conventional metallic
interfaces and surfaces. It is rather intriguing to observe that,
in spite of the complexity of the band structure arising from
complex hybridization among $s$, $p$, and $d$ orbitals, first
principle calculations do observe such a $k$-antisymmetric spin
splitting in the energy dispersion of interfacial states.
\cite{chantis,bilh,park} Although this spin splitting is more
subtle that the simple Rashba model depicted in Eq.
(\ref{eq:hamitonian}), it tends to confirm the phenomenological
intuition of Rashba \cite{rashba-soi} at metallic interfaces.

\subsection{Comparison between SHE torque and Rashba torque}
At this stage, it is interesting to compare the parameter
dependence of the in-plane torque $T_{\para}$ [in Eq.
(\ref{eq:torque-magnitude-weak-alp})] and the torque generated by
a spin-Hall effect \cite{she} in the bulk of a heavy metal
material such as Pt. In the latter case, the torque
$\bm{T}^{(\rm{SH})}$ exerted on the normal metal/ferromagnet
interface is obtained by projecting out the spin current
($\bm{j}^{(\rm{SH})}$ due to spin-Hall effect) that is transverse
to the magnetization direction.\cite{liu} In the bulk, the spin
current can be estimated using the ratio between spin-Hall
($\sg^{\tx{SH}}$) and longitudinal ($\sg_{xx}$) conductivities
(the so-called spin-Hall angle), \ie
\begin{equation}
j^{(\rm{SH})}=\frac{\sg^{\tx{SH}}}{\sg_{xx}}j_{e}.
\end{equation}
A perturbation calculation using the second-order Born
approximation gives rise to a spin current; thus the torque with a
magnitude given by
\begin{equation}
T^{(\rm{SH})}=\frac{\eta_{so}m \gm}{2e\tau_{\rm{tr}}^{0}} j_{e}
\label{eq:spin-hall-current}
\end{equation}
where, in general, $\gm > 1$ is a dimensionless parameter taking
into account both side-jump ($\gm=1$) and skew-scattering
($\gm>1$) contributions to the spin-Hall effect.
\cite{taka-mae-2008} $\eta_{so}$ is the spin-orbit-coupling
parameter and $\tau_{\rm{tr}}^{0}$ is the transport relaxation
time due to bulk impurities, the same definitions as in
Ref.[\onlinecite{taka-mae-2008}] except here the definition of
spin current differs by a unit $1/(2e)$. Meanwhile, the magnitude
of the Rashba-induced in-plane torque, i.e., Eq.
(\ref{eq:torque-magnitude-weak-alp}), can be simplified to, since
$\Dlt_{xc}\tau\ll 1$,
\begin{equation}
T_{\para}\approx \frac{4\alp m}{\eps_{F}e\tau_{sf}}j_{e}.
\label{eq:torque-para-parameter}
\end{equation}
The spin-orbit-coupling parameter $\alp$ in our definition in Eq.
(\ref{eq:hamitonian}) has the unit of energy. Equations
(\ref{eq:torque-para-parameter}) and (\ref{eq:spin-hall-current})
actually show that the in-plane Rashba torque and the spin-Hall
torque have a very similar parameter dependence. Meanwhile, a
diffusive description of the bilayer system, consisting of
ferromagnet/heavy metal, has shown that both SHE torque and Rashba
torque adopt a similar form,
\begin{equation}
\bm{T}=T_{\para}\bm{m}\times(\hat{\bm{y}}\times\bm{m})
+T_{\perp}\hat{\bm{y}}\times\bm{m}.
\end{equation}
The similarity in the geometrical form of the two torques implies
that, in principle, they are able to induce the same type of
magnetic excitation.\cite{shevsrashba}

The complexity of the underlying physics of spin-orbit torque and
the geometrical similarity between spin-Hall-induced and Rashba
torque make it a challenge to distinguish between two possible
origins. Recent progress has been made towards a plausible
distinction between the bulk and interfacial origin of different
torque components by varying the bilayer thickness,
\cite{kimhayashi} decoupling the heavy metal from the ferromagnet
\cite{fan} or dusting the interface with impurities,\cite{ryu}
which has revealed additional complex behaviors that question
current models including the Rashba model.

\section{Conclusion}
\label{sec:conclusion} Using Keldysh technique, in the presence of
both magnetism and a Rashba spin-orbit coupling, we derive a spin
diffusion equation that provides a coherent description to the
diffusive spin dynamics. In particular, we have derived a general
analytical expression for the Rashba torque in the bulk of a
ferromagnetic metal layer in both weak and strong Rashba limits.
We find that the spin-orbit torque in general consists of not only
in-plane and out-of-plane components but also a complex angular
dependence, which we attribute to the anisotropic spin relaxation
induced by the D'yakonov-Perel mechanism.

In the presence of magnetization dynamics, we have demonstrated
that our spin diffusion equation is able to describe a wealth of
phenomena including spin pumping and magnetic damping. In
particular, these results are in agreement with the earlier ones
derived using other methods. We have discussed the common features
shared by the Rashba and SHE torques. We also expect that further
investigation involving structural modification of the system
shall provide a deeper knowledge on the interfacial spin-orbit
interaction as well as the current-induced magnetization switching
in a single ferromagnet.

\begin{acknowledgments}
We thank G. E. W. Bauer, H. -W. Lee, K. -J. Lee, J. Sinova, M. D.
Stiles, X. Waintal, and S. Zhang for numerous stimulating
discussions.
\end{acknowledgments}

\end{document}